\documentclass[english,letterpaper,aps,prl,twocolumn,showpacs]{revtex4}
\usepackage{lmodern}

\usepackage[T1]{fontenc}
\usepackage[latin1]{inputenc}
\usepackage{amstext}
\usepackage{graphicx}
\usepackage{amssymb}

\makeatletter
\@ifundefined{textcolor}{}
{%
 \definecolor{BLACK}{gray}{0}
 \definecolor{WHITE}{gray}{1}
 \definecolor{RED}{rgb}{1,0,0}
 \definecolor{GREEN}{rgb}{0,1,0}
 \definecolor{BLUE}{rgb}{0,0,1}
 \definecolor{CYAN}{cmyk}{1,0,0,0}
 \definecolor{MAGENTA}{cmyk}{0,1,0,0}
 \definecolor{YELLOW}{cmyk}{0,0,1,0}
 }

\usepackage{lmodern}

\makeatletter

\usepackage{epsfig}

\usepackage{dcolumn}

\usepackage{bm}

\newlength{\textwidthm}

\makeatother

\makeatother

\makeatother

\makeatother

\makeatother

\usepackage{babel}

\begin{document}

\title{Superconducting States in pseudo-Landau Levels of Strained Graphene}

\author{Bruno Uchoa$^{1}$ and Yafis Barlas$^{2}$}

\affiliation{$^{1}$ Department of Physics and Astronomy, University of Oklahoma,
Norman, OK 73069, USA}

\affiliation{$^{2}$ Department of Physics and Astronomy, University of California
at Riverside, Riverside, CA 92521, USA}

\date{\today}
\begin{abstract}
We describe the formation of superconducting states in graphene in
the presence of pseudo-Landau levels induced by strain, when time
reversal symmetry is preserved. We show that superconductivity in
strained graphene is quantum critical when the pseudo-Landau levels
are completely filled, whereas at partial fillings superconductivity
survives at weak coupling. In the weak coupling limit, the critical
temperature scales \emph{linearly} with the coupling strength and
shows a sequence of quantum critical points as a function of the filling
factor that can be accessed experimentally. We argue that superconductivity
can be induced by electron-phonon coupling and that the transition
temperature can be controlled with the amount of strain and with the
filling fraction of the Landau levels. 
\end{abstract}

\pacs{71.27.+a,73.20.Hb,75.30.Hx}

\maketitle
Graphene is a single atomic sheet of carbon with electronic excitations
that behave as massless Dirac quasiparticles \cite{Novo2,Antonio}.
In general, graphene seems to be insensitive to electronic many body
instabilities \cite{Kotov}, except in the quantum Hall regime \cite{Barlas,Goerbig},
where fractional quantum Hall states \cite{Bolotin,Xu,Ghahari} have
been observed. We claim that one promising route to induce intrinsic
superconductivity in graphene is to reconstruct the electronic density
of states (DOS) into a discrete spectrum of Landau levels (LLs) with
the application of strain fields. Current experiments observed the
spontaneous formation of LLs on top of graphene nanobubbles \cite{Levy},
in deformed artificial graphene lattices, formed by a honeycomb grid
of molecules sitting on a metallic surface \cite{Gomes} and in chemical
vapor deposition grown graphene \cite{Yeh}. In specific engineered
forms \cite{Guinea,Guinea2}, applied strain configurations in graphene
mimic the application of strong uniform magnetic fields that can be
as large as 300T \cite{Levy,Gomes,Yeh}, but produce no net magnetic
flux, preserving time reversal symmetry (TRS). Quantum Hall states
induced by pseudomagnetic fields have been conjectured to give rise
to topological order in strained graphene with spontaneously broken
TRS \cite{Ghaemi,Abanin}. 

In this letter, we describe the formation of intrasublattice TRS spin
singlet states, which occupy the LLs produced by elastic deformations
in graphene. Since the overall wavefunction is anti-symmetric, the
spin singlet wavefunctions are even under valley exchange, and are
robust against backscattering \cite{Anderson}, unlike the corresponding
triplet states, which break inversion symmetry \cite{Ghaemi}. We
show that at integer filling factors where the normal state becomes
incompressible due to Pauli blocking, the superconducting order parameter
has a quantum critical point at the mean field level, $\Delta\propto|x-x_{c}|^{1/2}$,
where $x_{c}$ is a LL dependent critical coupling, in contrast with
the unstrained case, which is quantum critical only at the neutrality
point \cite{Uchoa0,Uchoa,Zhao}. At partial filling of the LLs, we
show that the zero temperature gap $\Delta\propto x$ has a \emph{linear}
scaling with coupling and strain in the weak coupling limit $x\ll x_{c}$,
and can be orders of magnitude larger than in conventional weak coupling
superconductors, where $\Delta\propto\mbox{e}^{-1/x}$ is suppressed
exponentially. Near complete filling of the LLs, a sequence of quantum
critical points can be experimentally accessed by controlling the
filling factor in the \emph{weak} coupling regime, $x<x_{c}$, opening
a prospect for the observation of quantum criticality in graphene.
We identify experimental signatures for this state, and propose that
in the presence of substrates which screen Coulomb interactions at
length scales larger than the magnetic length, superconductivity can
be triggered by conventional electron-phonon coupling. 

In the continuum description of the problem, the low energy electronic
Hamiltonian of strained graphene is \begin{equation}
\mathcal{H}_{0}=\sum_{\mathbf{\sigma,\alpha}}\int\mbox{d}\mathbf{x}\,\Psi_{\sigma,\alpha}^{\dagger}(\mathbf{x})\left[v(-i\nabla+\alpha\mathbf{A}_{s})\cdot\vec{\sigma}_{\alpha}-\mu\right]\Psi_{\sigma,\alpha}(\mathbf{x})\label{eq:Ho}\end{equation}
where $\sigma=\uparrow,\downarrow$ is the spin index, $\alpha=\pm$
indexes the two valleys, $\vec{\sigma}_{\alpha}=(\alpha\sigma_{x},\sigma_{y})$
is a vector of Pauli matrices, $\mu$ is the chemical potential away
from half filling, $v=6$eV$\AA$ is the Fermi velocity, $\Psi_{\sigma}=(\psi_{a,+,},\psi_{b,+},\psi_{a,-},\psi_{b,-})_{\sigma}$
is a 4 component spinor in the ($a,b$) sublattice pseudospin and
in the two valleys and $\mathbf{A}_{s}$ is the pseudo vector potential,
which couples to the electrons as a magnetic field pointing in opposite
directions in the two different valleys, preserving time reversal
invariance. The in-plane components of the pseudo magnetic field are
described by $A_{x}=u_{xy}$ and $A_{y}=\frac{1}{2}(u_{xx}-u_{yy})$,
and correspond respectively to strain and shear, where $u_{ij}=\nabla_{j}u_{i}+\nabla_{i}u_{j}+\nabla_{i}u_{z}\nabla_{j}u_{z}$
is the strain tensor, with $\mathbf{u}=(u_{x},u_{y},u_{z})$ the deformation
vector of the lattice normalized by the lattice constant \cite{Guinea}.
Although we assume strain configurations which produce approximately
uniform pseudo-magnetic fields \cite{Guinea,Guinea2}, $B_{s}=\nabla\times\mathbf{A}_{s}$,
this restriction is \emph{not} required for a macroscopic superconducting
state to emerge \cite{Note0}. 

In the Landau gauge, where $\mathbf{A}_{s}=(-B_{s}y,0)$, with $B_{s}$
the pseudo magnetic field, the electronic wavefunction takes the form
$\Psi_{k,\sigma}(x,y)=\mbox{exp}(ikx)\,\Theta_{\sigma}(y)$, where
$\Theta_{\sigma}(y)$ is the eigenspinor of a 1D Hamiltonian. This
Hamiltonian can be expressed in terms of ladder operators of the 1D
harmonic oscillator, $a\equiv\left(\xi+\partial_{\xi}\right)/\sqrt{2},\, a^{\dagger}\equiv\left(\xi-\partial_{\xi}\right)/\sqrt{2},$
where $\xi\equiv\ell_{B}k-y/\ell_{B}$ is a dimensionless variable
related to the valley dependent guiding center $X=-k\ell_{B}^{2}$,
with $\ell_{B}=\sqrt{\hbar/eB_{s}}$ (restoring $\hbar$) the effective
magnetic length, and $e$ the electron charge. In what follows, we
define the valley dependent operator $\hat{\mathcal{D}}(\xi)$, \begin{equation}
v(-i\nabla+\mathbf{A}_{s})\cdot\vec{\sigma}=\sqrt{2}\frac{v}{\ell_{B}}\left(\begin{array}{cc}
0 & a\\
a^{\dagger} & 0\end{array}\right)\equiv\hat{\mathcal{D}}(\xi),\label{eq:D}\end{equation}
which takes the form $-v(-i\nabla-\mathbf{A}_{s})\cdot\vec{\sigma}=-\hat{\mathcal{D}}(\bar{\xi})$
in the opposite valley, with $\bar{\xi}=\ell_{B}k+y/\ell_{B}$. 

In the presence of an effective attractive potential $U$ that stabilizes
the superconducting state, the Bogoliubov-deGennes (BdG) Hamiltonian
is $\mathcal{H}_{\textrm{BG}}=\int\mbox{d}\mathbf{x}\,\Phi^{\dagger}(\mathbf{x)}\hat{\mathcal{H}}_{BG}\Phi(\mathbf{x})$,

\begin{equation}
\hat{\mathcal{H}}_{\textrm{BG}}=\left(\begin{array}{cc}
\hat{\mathcal{H}}_{0}(\mathbf{A}_{s}) & \hat{\Delta}\\
\hat{\Delta}^{*} & -\mathcal{T}\hat{\mathcal{H}}_{0}(\mathbf{A}_{s})\mathcal{T}^{-1}\end{array}\right),\label{eq:Hbg}\end{equation}
where $\hat{\mathcal{H}}(\mathbf{A}_{s})=\hat{\mathcal{D}}(\xi)\otimes\nu^{+}-\hat{\mathcal{D}}(\bar{\xi})\otimes\nu^{-}-\mu\sigma_{0}\nu_{0},$
is the normal state Hamiltonian of strained graphene written in valley
and sublattice spaces, $\nu^{\pm}=(\nu_{0}\pm\nu_{z})/2$ are projectors
in the $\pm$ valley spaces, with Pauli matrices $\nu_{i}\,(i=x,y,z)$,
and $\Phi=(\Psi_{k,\uparrow},\Psi_{-k,\downarrow}^{\dagger})$ is
the $8$ component spinor in the Nambu space, with Pauli matrices
$\tau_{i}$ ($i=x,y,z)$. The off diagonal term, $\hat{\Delta}$,
is a pairing matrix that describes the formation of Cooper pairs $\Delta^{2}=U\mbox{tr}\langle\Psi_{k,\sigma}^{\dagger}\hat{\Delta}\Psi_{-k,-\sigma}\rangle$.
In strained graphene, the time reversal symmetry operation $\mathcal{T}\hat{\mathcal{H}}_{0}(\mathbf{A}_{s})\mathcal{T}^{-1}$
leaves the Hamiltonian invariant under an additional exchange between
valleys, in contrast with the case of conventional magnetic fields,
which explicitly break TRS \cite{Tesanovic}. 

In the intra-sublattice $s$\textbf{\emph{-}}wave pairing state, which
corresponds to the pairing matrix \textbf{\emph{$\hat{\Delta}=\Delta\sigma_{0}\nu_{x}$,
}}the eigenvector problem $\hat{\mathcal{H}}_{\textrm{BG}}\Phi(x,\xi)=E\Phi(x,\xi)$
can be solved by decomposing Hamiltonian (\ref{eq:Hbg}) into two
equivalent copies of $4\times4$ BdG Hamiltonians in pseudospin and
Nambu spaces. In the reduced Nambu basis $\bar{\Phi}=(\Psi_{\uparrow,+},\Psi_{\downarrow,-}^{\dagger})$,
\begin{equation}
\bar{\mathcal{H}}_{BG}=\left(\begin{array}{cc}
\hat{\mathcal{D}}(\xi)-\mu & \Delta\\
\Delta^{*} & -\hat{\mathcal{D}}(\xi)+\mu\end{array}\right).\label{eq:Hbg2}\end{equation}
Fixing the gauge of the gap $\Delta$ to be real, the eigenvalue problem
\emph{$(E-\bar{\mathcal{H}}_{BG})\bar{\Phi}=0$} is equivalent to
$\mathcal{M}\bar{\Phi}\equiv(E+\bar{\mathcal{H}}_{BG}^{\prime})(E-\bar{\mathcal{H}}_{BG})\bar{\Phi}=0$
where $\bar{\mathcal{H}}_{BG}^{\prime}\equiv\mathcal{C}\bar{\mathcal{H}}_{BG}\mathcal{C}^{-1}=(\hat{\mathcal{D}}+\mu)\otimes\tau_{3}+\Delta\tau_{1}$
is the charge conjugated BdG Hamiltonian ($\mu\to-\mu$). When the
matrix $\mathcal{M}$ is applied in a proper basis, $\mathcal{M}$
can be cast in the form $\mathcal{M}\bar{\Phi}=(E+\mathcal{H}_{+})(E-\mathcal{H}_{-})\bar{\Phi}$,
with $\mathcal{H}_{\pm}=(s\omega_{c}\sqrt{|N|}\pm\mu)\tau_{3}+\Delta\tau_{1},$
where $N$ is the index of the Landau levels, $s(N)\equiv\mbox{sgn}(N)$
accounts for the two branches of LLs in the conduction and valence
bands, and $\omega_{c}=\sqrt{2}v/\ell_{B}$. $\mathcal{H}_{-}$ is
equivalent to Hamiltonian (\ref{eq:Hbg2}) and gives the energy spectrum
\begin{equation}
\pm E_{N}=\pm[(s\omega_{c}\sqrt{|N|}-\mu)^{2}+\Delta^{2}]^{1/2},\label{eq:EN}\end{equation}
with eigenstates given by \begin{equation}
\Psi_{\pm,\sigma,\alpha}^{(N)}(\xi)=\beta_{\pm,\sigma,\alpha}^{N}\left(\begin{array}{c}
\phi_{|N|-1}(\xi)\\
s\phi_{|N|}(\xi)\end{array}\right)\mbox{e}^{ikx},\label{eq:Psi2}\end{equation}
for the states $(\sigma,\alpha)=(\uparrow,-)$ and ($\downarrow,+$),
where $\beta_{\pm,\uparrow,-}^{N}=1/\sqrt{2}[1\pm(s\omega_{c}\sqrt{|N|}-\mu)/E_{N}]^{1/2}$,
and $\beta_{\pm,\downarrow,+}^{N}=\mp1/\sqrt{2}[1\mp(s\omega_{c}\sqrt{|N|}-\mu)/E_{N}]^{1/2}$.
$\phi_{N}(\xi)$ denotes conventional LL wavefunctions, with $\phi_{-1}(\xi)=0$.
In the zero LL, the Cooper pairs occupy only one sublattice, explicitly
breaking the $\mathbb{Z}_{2}$ sublattice symmetry of graphene. As
anticipated, the BdG quasiparticle spectrum is discrete and can be
indexed by the LL index $N$. The superconducting ground state is
given by \cite{Note}\begin{equation}
|\Psi_{0}\rangle=\prod_{N,X}\left(u_{N}+v_{N}c_{N,X,\uparrow,-}^{\dagger}c_{N,-X,\downarrow,+}^{\dagger}\right)|0\rangle,\label{eq:Psi2-1}\end{equation}
where $u_{N}=\beta_{+,\uparrow,-}^{N}$, $v_{N}=-\beta_{+,\downarrow,+}^{N}$
and $c_{N,X,\sigma,\alpha}^{\dagger}$ are fermionic creation operators
of the relativistic LLs. This wavefunction describes intrasublattice
pairing across opposite valleys, within the same LL. As in usual spin
singlet superconductivity, the essence of Cooper phenomenon, that
TRS states can pair up, is preserved by the discrete spectrum of LLs. 

\begin{figure}
\begin{centering}
\includegraphics[scale=0.28]{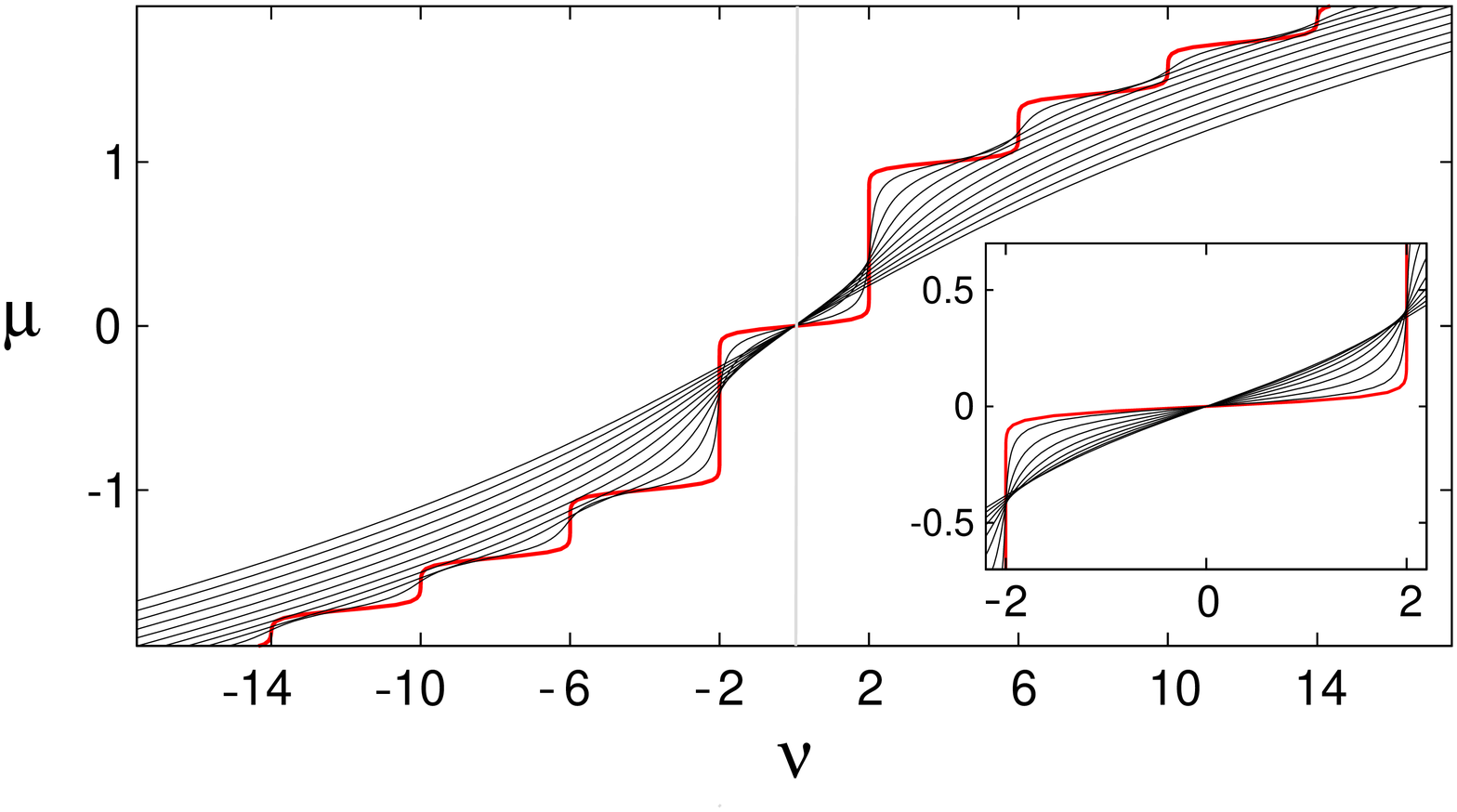}\vspace{-0.3cm}
\par\end{centering}

\caption{{\small Dependence of the chemical potential $\mu$ in units of $\omega_{c}$
with the filling factor $\nu$. Red (light) curve: normal state $\Delta=0$.
Solid black curves: superconducting state, with $\Delta/\omega_{c}$
ranging from $0$ (red line) to 1. All curves are plotted at $T/\omega_{c}=0.02$.
Inset: detail of the zero LL $(n=0$). The red curve is analytically
described by Eq. (\ref{eq:mu}), while solid ones at $\Delta/\omega_{c}\ll1$
are described by Eq. (\ref{eq:muSC}) in the $T\to0$ limit. }}
\end{figure}

The discontinuous behavior of the chemical potential with the pseudomagnetic
field and filling factor can be calculated by fixing the total number
of particles in the system. Although the ground state wave function
does not conserve the number of particles, the distribution is sharply
peaked around the average $\mathcal{N}$ in the thermodynamic limit
\cite{Tinkham}, $\mathcal{N}=gN_{\phi}\sum_{N=-\infty}^{\infty}\left[u_{N}^{2}f(E_{N})+v_{N}^{2}f(-E_{N})\right],$
where $f(E)=(1+\mbox{e}^{E/T})^{-1}$ is the Fermi distribution, $g=4$
is the valley and spin degeneracy, and $N_{\phi}=A/(2\pi\ell_{B}^{2})$
is the number of flux quanta for a total area $A$, which sets the
LLs degeneracy. In the low temperature and weak coupling regime $T,\Delta\ll v/\ell_{B}$,
where the deep energy states $N<n$ are fully occupied, with $n$
the highest occupied LL, the constraint becomes \begin{equation}
2(\nu/g-n)=-[(s\omega_{c}\sqrt{|n|}-\mu)/E_{n}]\tanh[E_{n}/(2T)],\label{eq:nu3}\end{equation}
where $\nu=\mathcal{N}/N_{\phi}-g(N_{\Lambda}+1/2)$ is the filling
factor, and $N_{\Lambda}=(D/\omega_{c})^{2}>0$ is an ultraviolet
cutoff that regularizes the number of negative energy states, where
$D\sim6$eV is the bandwidth. In particular, at $T=0$, the chemical
potential \begin{equation}
\mu(0,\nu)=s\omega_{c}\sqrt{|n|}+\frac{\Delta(\nu-gn)}{\sqrt{[g(n+\frac{1}{2})-\nu][\nu-g(n-\frac{1}{2})]}},\label{eq:muSC}\end{equation}
remains pinned to the $n$-th LL when half filled ($\nu=gn)$ for
small $\Delta$, and shows a power law divergence when the highest
occupied LL is completely filled, at integer fillings $\nu=g(n\pm1/2),$
indicating an incompressibility due to Pauli blocking. In the opposite
regime, when $T,\Delta\gtrsim v/\ell_{B}$, the system crosses over
to the usual Fermi liquid behavior when the electrons have multiple
transitions between different LLs. In the normal state ($\Delta=0$),
as expected, the chemical potential \begin{equation}
\mu(T,\nu)=s\omega_{c}\sqrt{|n|}+T\ln\left[\frac{\nu-g(n-\frac{1}{2})}{g(n+\frac{1}{2})-\nu}\right],\label{eq:mu}\end{equation}
has a logarithmic divergence at integer filling. Eq. (\ref{eq:muSC})
and (\ref{eq:mu}) describe analytically the numerical curves shown
in the inset of Fig. 1 when $\Delta/\omega_{c}\ll1$. 

When all LLs are taken into account, these divergences are regularized,
as shown in Fig. 1, leading to a sequence of jumps. At integer fillings
$\nu_{\mathrm{I}}(n)=g(n+\frac{1}{2})$, the chemical potential for
the normal state does not diverge but sits half way between the LLs
in the the zero temperature limit, $h(n)\equiv\mu[0,\nu_{\mathrm{I}}(n)]/\omega_{c}=[s(n+1)\sqrt{|n+1|}+s(n)\sqrt{|n|}]/2$.
The red (light) curve in Fig. 1 describes the $\Delta=0$ case at
fixed temperature, while the black lines represent the superconducting
case for fixed values of the gap. Unlike in conventional superconductors,
in TRS LLs the chemical potential $\mu$ has a strong dependence with
the gap, which must be accounted self consistently into the equation
of state. 

At the mean field level, the free energy of the superconducting state
is $F=-TgN_{\phi}\sum_{\gamma=\pm}\sum_{N=-\infty}^{\infty}\ln(1+\mbox{e}^{-\gamma E_{N}/T})-\bar{A}|\Delta|^{2}/U$,
where $\bar{A}$ is the total area normalized by the size of the unit
cell. Minimization of the free energy gives the gap equation\begin{equation}
1=-(U/2)g\bar{N}_{\phi}\sum_{N=-\infty}^{\infty}\tanh\left[E_{N}(T,\nu)/(2T)\right]/E_{N},\label{eq:gap}\end{equation}
where $\bar{N}_{\phi}=3\sqrt{3}a^{2}/(4\pi\ell_{B}^{2})$ is the number
of flux quanta per unit cell, with $a=1.42\mbox{\AA}$ the lattice
spacing. Defining $x\equiv|U|g\bar{N}_{\phi}/\omega_{c}\propto|U|/\ell_{B}$
as the dimensionless coupling parameter that controls the strength
of interactions and strain, at half filling $(\nu=0)$, the zero temperature
gap in the weak coupling regime $T_{c}\ll v/\ell_{B}$ is \begin{equation}
\Delta^{(0)}(0)=\left[v\, x/(\sqrt{2}\ell_{B})\right]/[1-\zeta_{A}\!\left(1/2\right)x],\label{eq:Tc}\end{equation}
where $\zeta_{A}(\frac{1}{2})=\sum_{N=1}^{N_{\Lambda}}1/\sqrt{N}$
is the zeta function regularized by an ultraviolet cut-off. The ratio
between the critical temperature and the zero temperature gap at half
filling is a universal number, $2T_{c}=\Delta^{(0)}(0)$. In the weak
coupling limit, $T_{c}\sim\sqrt{2}vx/\ell_{B}\propto B_{s}$ has a
linear scaling with the coupling and with the amount of strain \cite{Kopnin}.
This scaling contrasts with the case of conventional weak coupling
superconductors, where $T_{c}\propto\mbox{exp}(-1/x)$ decreases exponentially
with the effective coupling. As the coupling $x$ becomes larger,
the system eventually crosses over to the strong coupling regime,
when $T_{c}\gtrsim v/\ell_{B}$, as shown in Fig. 2. In the critical
regime, when $\Delta/T_{c}\ll1$, the gap at $\nu=0$ is given by
\begin{equation}
\Delta^{(0)}(T)=2^{\frac{1}{4}}(v/\ell_{B})^{\frac{3}{2}}\sqrt{1-T/T_{c}},/[\zeta\left(3/2\right)T_{c}]^{\frac{1}{2}},\label{eq:Delta}\end{equation}
at weak coupling, and scales with $\Delta^{(0)}\propto B_{s}^{3/4}$,
where $\zeta(3/2)\approx2.61$ is a zeta function. In Fig. 2a, we
show the dependence of the critical temperature with the coupling
$x$ for different filling factors. The red curve is the phase transition
for $\nu=2$, which is quantum critical. 

\begin{figure}
\begin{centering}
\includegraphics[scale=0.33]{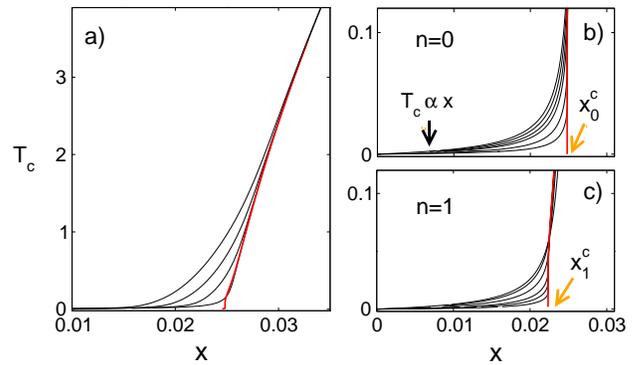}
\par\end{centering}

\caption{{\small Critical temperature $T_{c}/\omega_{c}$ vs. the dimensionless
coupling strength $x\equiv|U|gN_{\phi}\ell_{B}/(\sqrt{2}v)$. a) red}
{\small (light) curve: $\nu=2$, where the transition is quantum critical
below $x_{0}^{c}\approx0.025$. Solid black curves: $\nu=0,\,8,\,24$
and $\nu=40$, from right to left.} b) {\small $\nu=0,\,1.2,\,1.6,\,1.98$,
$1.9998$ and 2, from left to right. c) $\nu=4,\,5.2,\,5.6\,5.98,\,5.9998$
and $6$, from left to right. At partial filling of the LL, $T_{c}\propto x$
in the $x\to0$ limit.}}
\end{figure}

At integer filling factors $\nu_{\mathrm{I}}(n)=g(n+\frac{1}{2})$,
when the highest LL is completely filled, the chemical potential of
the normal state sits half way between the LLs at zero temperature.
The normal state becomes incompressible, and the emergence of superconductivity
requires a quantum critical coupling, which allows transitions between
different LLs. At those integer fillings, the zero temperature gap
is \begin{equation}
\Delta^{(\nu_{\mathrm{I}})}(0)=2\sqrt{2\gamma_{n}}(v/\ell_{B})\sqrt{x/x_{n}^{c}-1},\label{eq:deltan}\end{equation}
where \begin{equation}
x_{n}^{c}=2/\sum_{N=-N_{\Lambda}}^{N_{\Lambda}}|s\sqrt{|N|}-h(n)|^{-1}\label{eq:xc}\end{equation}
is the quantum critical coupling of the $\nu=g(n+\frac{1}{2})$ state,
when the $n$-th LL is completely filled, $\gamma_{n}^{-1}=\sum_{N=-\infty}^{\infty}|s\sqrt{|N|}-h(n)|^{-3}$
is a constant and $h(n)$ was defined was defined below Eq. (\ref{eq:mu}).
For a magnetic length of $\ell{}_{B}\sim30\mbox{\AA}$\emph{, }which
corresponds to a pseudomagmetic field $B_{s}\sim100$T, we have $N_{\Lambda}\sim400$.
At $n=0$, the critical coupling is $x_{0}^{c}\approx0.025$, as depicted
by the arrow in Fig. 2b for the red curve, crossing over to a smooth
transition for partial filling factors, as shown in the black curves
in the same panel, where $0\leq\nu<2$. For $n=1$, the critical coupling
drops to $x_{1}^{c}\approx0.022$ (Fig. 2c), with the solid curves
in the same panel indicating a crossover to partial filling factors
in the range $4\leq\nu<6$. For higher LL, $x_{n}^{c}$ decreases
further as $n$ becomes large. 

\begin{figure}[t]
\begin{centering}
\includegraphics[scale=0.34]{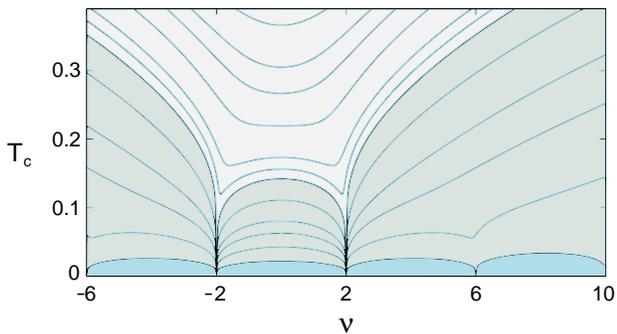}
\par\end{centering}

\caption{{\small Phase diagram $T_{c}/\omega_{c}$ as a function of the filling
factor $\nu$ for different coupling strengths, $x$ (solid curves).
Blue (dark) region: $x<x_{1}^{c}\approx0.022$. Grey region: $x_{1}^{c}<x<x_{0}^{c}\approx0.025$.
Light region: $x>x_{0}^{c}$. The critical temperature drops to zero
at integer filling factors $\nu_{\mathrm{I}}(n)=g(n+\frac{1}{2})$
whenever $x<x_{n}^{c}$, where $x_{n}^{c}$ is the critical coupling
of the $\nu_{\mathrm{I}}(n)$ state. }}
\end{figure}

For fixed coupling strength, the critical temperature evolves with
$\nu$ as a series of lobes and drops to zero at integer filling factors
$\nu_{\mathrm{I}}(n)=g(n+\frac{1}{2})$, whenever $x<x_{n}^{c}$,
as shown in Fig. 3. The solid blue area in Fig. 3 depicts the region
$x<x_{1}^{c}$, which is quantum critical for the first two Landau
levels, while the gray region $x_{1}^{c}<x<x_{0}^{c}$ is quantum
critical in the zero LL only. When $x$ grows larger than $x_{n}^{c}$,
the system undergoes a quantum phase transition at the filling factor
$\nu=\nu_{I}(n)$, when superconductivity emerges. For $\nu\sim\nu_{\mathrm{I}}(n)$,
the critical temperature scales as \begin{equation}
T_{c}(x,\nu)\propto|\nu-\nu_{\mathrm{I}}(n)|^{\delta}\label{eq:Tcnu}\end{equation}
 for $x<x_{n}^{c}$, with possible logarithmic behavior, where $\delta\sim0.2$
is the exponent numerically extracted for the $\nu=\pm2$ states.
This behavior may lead to the experimental observation of quantum
criticality in graphene by controlling the filling factor of the LL
in the \emph{weak} coupling limit $x\ll x_{n}^{c}$. In two dimensions,
the mean-field critical temperature $T_{c}$ sets the onset of Cooper
pair formation, while phase coherence is lost above the Kosterlitz-Thouless
(KT) transition temperature $T_{KT}<T_{c}$ \cite{Loktev}, where
pairs of vortices and anti-vortices unbind. Phase fluctuations will
likely be relevant for transport and will be discussed elsewhere.

Besides transport measurements in \emph{long} graphene junctions,
where a supercurrent is expected to flow along the edges \cite{Covaci},
one experimental signature of superconductivity in strained graphene
is the specific heat at fixed volume, $C_{V}=-T(\partial^{2}F/\partial T^{2})_{V}$.
At low temperature, $T\ll T_{c}\ll v/\ell_{B}$, the specific heat
of the $\nu=0$ state is $C_{V}(T)=2gN_{\phi}\Delta^{2}(0)\mbox{e}^{-\Delta(0)/T}/T^{2}$.
In the quantum limit, $\omega_{c}/T_{c}\gg1$, the specific heat jump
at the phase transition normalized by the specific heat in the normal
side for $\nu=0$ is $\Delta C/C_{n}=\omega_{c}\mbox{e}^{\omega_{c}/T_{c}}/[16\zeta(3/2)T_{c}]$
which is \emph{non-universal}. In the weak coupling regime $x\ll x_{0}^{c}$,
where $T_{c}\sim x\omega_{c}/4$, \begin{equation}
(\Delta C/C_{n})(x)=\mbox{e}^{4/x}/[4x\,\zeta(3/2)]\label{eq:deltaC}\end{equation}
becomes exponentially large as $T_{c}$ drops to zero. This feature
is a signature of this state, and contrasts both with the specific
heat jump expected for Dirac fermion superconductivity in unstrained
graphene at half filling ($\Delta C/C_{n}\approx0.35$) \cite{Uchoa0}
and in weak coupling superconductors in general ($\Delta C/C_{n}\approx1.43$)\cite{Tinkham},
which are universal constants. 

Although strong Coulomb interactions inhibit superconductivity and
can give rise to incompressible states at fractional filling factors
\cite{Barlas}, a condensate can be induced by phonons in the presence
of substrates that screen the electronic repulsion at length scales
larger than $\ell_{B}$. The analysis of scanning tunneling spectroscopy
experiments \cite{Miller} in graphene on SiC for magnetic fields
around 5T, when the LLs are well defined, indicate that the effective
momentum independent electron phonon vertex in graphene is $g_{0}\sim0.1$eV
for an Einstein phonon mode at the typical frequency $\omega_{{\rm ph}}\sim0.2$eV
\cite{Miller,Pound}. This mode alone ($E_{2g}$ phonon) leads to
an effective attraction $U\sim-2g_{\textrm{0}}^{2}/\omega_{{\rm ph}}\approx-0.1$
eV. For a magnetic length of 20$\mbox{\AA}$, which corresponds to
$2\times10^{-3}$ flux quanta per unit cell, a net attractive coupling
of that order results in a dimensionless coupling $x\sim0.003$ and
a critical temperature $T_{c}\sim8$K at the $\nu=0$ state. 

In the current experiments where pseudomagentic fields of 300 T were
observed on the surface of graphene nanobubbles with $10$nm in size
each \cite{Levy}, a macroscopic superconducting state can emerge
when the average spacing among the nanobubbles $b\sim40$nm is shorter
than the coherence length $\xi\sim v/(\pi\Delta)$\cite{Note0}. For
instance, a superconducting gap of $\Delta\sim$1 meV ($T_{c}\sim10$K),
corresponds to a coherence length $\xi\sim200$nm, which sets the
length the Cooper pairs created on top of the bubbles can travel coherently
in the normal unstrained regions. 

A significant enhancement of the electron-phonon coupling, and as
a result $T_{c}$, can be achieved for instance by coating graphene
with ionic crystals and alkaline metals such as K, which is known
to form a stable crystal on top of graphene \cite{Caragiu,Uchoa3,Gruneis}.
This mechanism can lead to measurable transition temperatures in the
regime where the broadening of the highest occupied LL due to disorder
effects is small compared to the level spacing. Our analysis shows
that the spin singlet states are robust, and present a sequence of
quantum critical points, which can be experimentally accessed by tuning
the filling factor of the LLs in the weak coupling limit of the problem. 

The authors acknowledge M. Fogler, F. Guinea, K. Mullen, A. Jaefari,
C. Bolech, N. Shah, C. Varma and V. Aji for discussions. BU acknowledges
financial support from University of Oklahoma during the summer.

\end{document}